\documentclass[11pt]{article}

\usepackage[utf8]{inputenc}
\usepackage[T1]{fontenc}
\usepackage{lmodern}
\usepackage{amsmath,amssymb,amsthm}
\usepackage{booktabs}
\usepackage{graphicx}
\usepackage{xcolor}
\usepackage{hyperref}
\usepackage{enumitem}
\usepackage{geometry}
\usepackage{algorithm}
\usepackage{algpseudocode}
\hypersetup{
  colorlinks=true,
  linkcolor=blue,
  citecolor=blue,
  urlcolor=blue,
}

\newcommand{\MLKEM}{\textsf{ML-KEM}}
\newcommand{\Kyber}{\textsf{CRYSTALS-Kyber}}
\newcommand{\KyFrog}{\textsf{KyFrog}}
\newcommand{\algo}[1]{\textsf{#1}}

\title{%
  KyFrog:\\
  A High-Security LWE-Based KEM Inspired by ML-KEM
}

\author{%
  V\'ictor Duarte Melo%
  \thanks{Independent Researcher, \texttt{victormeloasm@gmail.com}.} \and
  William J.\ Buchanan%
  \thanks{Edinburgh Napier University, \texttt{b.buchanan@napier.ac.uk}.}
}

\date{\today}

\begin{document}
\maketitle

\begin{abstract}
NIST's Module-Lattice-Based Key-Encapsulation Mechanism (\MLKEM{}) standard, derived from the \Kyber{} family~\cite{kyber,fips203}, is the current reference design for post-quantum key encapsulation.
Its parameter sets ML-KEM-512/768/1024 target NIST security categories 1, 3 and 5 with relatively small public keys and ciphertexts.
In this work, we explore a complementary operating point in the design space: an aggressively conservative Learning-with-Errors (LWE) key-encapsulation mechanism called \KyFrog{}.
KyFrog uses a larger dimension ($n=1024$) and a small prime modulus $q=1103$, together with narrow error distributions with standard deviations $\sigma_s = \sigma_e = 1.4$, to target approximately $2^{325}$ classical and quantum security against state-of-the-art lattice attacks under standard cost models, as estimated using the \emph{Lattice Estimator}~\cite{albrecht-lattice-estimator}.
The price paid for this security margin is an extremely large KEM ciphertext (about 0.5\,MiB), while public and secret keys remain in the same ballpark as ML-KEM.
We describe the design rationale, parameter search methodology, and implementation details of KyFrog, and we compare its asymptotic security and concrete parameter sizes with the \MLKEM{} standard.
All code and data for this work are released as free and open-source software, with the full C++23 implementation and experimental scripts available at \url{https://github.com/victormeloasm/kyfrog}~\cite{kyfrog-github}.
\end{abstract}

\section{Introduction}
The migration from classical public-key cryptography to post-quantum cryptography (PQC) is one of the most significant transitions in the history of applied cryptography.
Once large-scale quantum computers become available, widely deployed schemes such as RSA and elliptic-curve cryptography (ECC) will be vulnerable to Shor's algorithm.
New primitives must therefore be deployed \emph{before} quantum adversaries become realistic, in order to protect long-lived secrets and stored ciphertexts.

Among the families proposed for PQC, lattice-based cryptography---and in particular Learning-with-Errors (LWE) and Module-LWE (MLWE) constructions---has emerged as a leading candidate due to its strong worst-case-to-average-case reductions and its competitive performance.
NIST's selection and standardization of \Kyber{} as the basis for the \MLKEM{} standard~\cite{kyber,fips203} confirms the central role of lattice-based key-encapsulation mechanisms (KEMs) in the post-quantum ecosystem.

While the \MLKEM{} parameter sets provide well-balanced trade-offs between security and efficiency, it is conceptually useful to explore more extreme points in the design space.
In some scenarios, bandwidth is relatively cheap but long-term security margins are paramount: examples include key escrow, cold-wallet key wrapping, or archival protection of high-value secrets.
In such settings, it may be acceptable to pay a large price in ciphertext size to gain a more generous security buffer against future cryptanalytic advances.

This paper introduces \KyFrog{}, an LWE-based KEM designed explicitly for this ``paranoid'' regime.
Rather than optimizing for compact ciphertexts, KyFrog maximizes estimated lattice-attack cost under conservative models, using a simple and transparent parameter set with $n=1024$ and a prime modulus $q=1103$.
A separate tool, \emph{KyFrog Hunter}, is used to explore and validate candidate parameter sets, integrating the \textsf{lattice-estimator} in an automated pipeline~\cite{albrecht-lattice-estimator}.

\subsection{Motivation for a ``paranoid'' KEM}
Most deployed and standardized KEMs, including \MLKEM{}, aim to minimize bandwidth and computational cost for a given target security category.
However, there exist niche but important use cases where the dominant constraint is \emph{not} bandwidth but rather the desire for a large security margin and a very small probability of future breakage.
Examples include the protection of master secrets in long-term hardware security modules, key escrow for high-value forensic material, and the encryption of regulatory archives that must remain confidential for several decades.

In such settings, an ``overkill'' parameter set that sacrifices ciphertext size for a dramatically increased estimated work factor may be reasonable.
KyFrog occupies this design point: it is not intended as a drop-in competitor to ML-KEM in bandwidth-constrained environments, but rather as a conceptual and practical demonstration of how far we can push LWE parameters within a simple, auditable design.

\subsection{Overview of the paper}
Section~2 reviews the LWE and MLWE problems, the \Kyber{}/\MLKEM{} design, and the role of the \textsf{lattice-estimator} in concrete security estimates.
Section~3 outlines KyFrog's design goals and threat model.
Section~4 describes the core construction and encoding of KyFrog, and Section~5 provides a formal specification of the key-generation, encapsulation and decapsulation algorithms.
Section~6 focuses on correctness and decryption failure probabilities.
Section~7 details the KyFrog Hunter parameter search.
Section~8 discusses security considerations and attack surfaces.
Section~9 covers implementation aspects, including randomness, symmetric encryption and constant-time coding.
Section~10 describes our benchmarking methodology, while Section~11 explores potential use cases and deployment considerations.
Section~12 discusses limitations and open questions, and Section~13 concludes with directions for future work.
Appendices~\ref{app:keygen-instance} and~\ref{app:hunter-stats} document concrete key-generation reports and KyFrog Hunter run statistics.

\section{Background}
\subsection{The Learning-with-Errors Problem}
The Learning-with-Errors (LWE) problem is parameterized by a dimension $n$, a modulus $q$, and a noise distribution $\chi$ over $\mathbb{Z}_q$.
An LWE sample consists of a pair $(\mathbf{a}, b) \in \mathbb{Z}_q^n \times \mathbb{Z}_q$ where
\begin{align*}
  \mathbf{a} &\leftarrow \mathbb{Z}_q^n,\\
  e &\leftarrow \chi,\\
  b &= \langle \mathbf{a}, \mathbf{s} \rangle + e \pmod{q},
\end{align*}
for some secret vector $\mathbf{s} \in \mathbb{Z}_q^n$.
The decisional LWE problem asks to distinguish such samples from uniformly random pairs $(\mathbf{a}, b)$ over $\mathbb{Z}_q^n \times \mathbb{Z}_q$.
The hardness of LWE is supported by reductions from worst-case lattice problems such as GapSVP and SIVP.

\subsection{Module-LWE and structured lattices}
Module-LWE (MLWE) generalizes LWE by working over modules of dimension $k$ in a polynomial ring $R_q = \mathbb{Z}_q[X]/(X^n+1)$, which allows compact representations and efficient Number-Theoretic Transform (NTT) implementations.
\Kyber{} and \MLKEM{} are defined in terms of MLWE and exploit this structure for performance~\cite{kyber,fips203}.
The underlying lattices inherit a great deal of algebraic structure, which is both a blessing (fast arithmetic and compact encodings) and a source of caution (potential structural attacks).
To date, no practical attacks exploiting the ring structure of MLWE-based schemes such as \Kyber{} have been found, but conservative designs sometimes prefer to fall back to plain LWE or to module dimensions $k$ as small as possible.

\subsection{CRYSTALS-Kyber and ML-KEM}
\Kyber{} is a CCA-secure KEM whose security is based on MLWE.
It was selected as the primary KEM in NIST's PQC process and later standardized as \MLKEM{}~\cite{kyber,fips203}.
The three standardized parameter sets, ML-KEM-512, ML-KEM-768 and ML-KEM-1024, share $n=256$ and $q=3329$ but differ in the module rank $k$ and noise parameters.
Their targeted security levels roughly correspond to AES-128, AES-192, and AES-256, respectively.

Each parameter set defines the sizes of the public key (encapsulation key), secret key (decapsulation key), and ciphertext.
For instance, ML-KEM-1024 uses a 1568-byte public key, a 3168-byte secret key, and a 1568-byte ciphertext, with tight bandwidth requirements and excellent performance on modern CPUs and embedded platforms.

\subsection{Security notions for KEMs}
KyFrog aims to achieve standard indistinguishability under adaptive chosen-ciphertext attack (IND-CCA2) in the KEM setting.
At an informal level, IND-CCA2 security requires that an adversary who is allowed to obtain decapsulated keys for arbitrary ciphertexts of her choice (except for a single challenge ciphertext) cannot distinguish the real session key associated with the challenge from a uniformly random key.
The \Kyber{} and \MLKEM{} constructions achieve IND-CCA2 security by applying a Fujisaki--Okamoto-style transform~\cite{fujisaki-okamoto} to an underlying IND-CPA-secure public-key encryption scheme~\cite{kyber,fips203}.

KyFrog follows the same paradigm: we first define a simple LWE-based public-key encryption scheme, prove or inherit its IND-CPA security from the hardness of LWE, and then apply an FO-style KEM transform.
The transform uses hashing, re-encryption and constant-time checks to prevent decryption-oracle attacks that attempt to distinguish valid and invalid ciphertexts based on timing or error messages.

\subsection{Security estimation for lattice KEMs}
Concrete security estimates for lattice-based schemes are typically obtained by modeling the cost of lattice reduction and secret-key recovery attacks via tools such as the \textsf{lattice-estimator}~\cite{albrecht-lattice-estimator}.
These tools consider families of attacks (e.g., primal, dual, hybrid) and assign a bit-security estimate under classical and quantum cost models.

While estimates are inherently heuristic, they provide a common language to compare parameter sets.
In this work, KyFrog is tuned to reach an estimated classical and quantum security of approximately 325.3 bits, well above ML-KEM-1024.
We treat these numbers as guidance rather than proofs: the goal is to illustrate the trade-off between bandwidth and a generous security margin.

\subsection{Related work on conservative parameter choices}
Several LWE and MLWE-based schemes have explored conservative parameter choices or unstructured lattices to increase confidence in security claims.
FrodoKEM, for example, works directly with matrix LWE over $\mathbb{Z}_q$ without using ring structure, at the cost of larger key and ciphertext sizes~\cite{frodokem}.
Other works study systematic parameter tuning using tools such as the \textsf{lattice-estimator}, exploring the impact of $n$, $q$ and noise parameters on the estimated cost of attacks.
KyFrog can be seen as a lightweight, implementation-focused contribution to this line of work, emphasizing a transparent design and a fully reproducible search for a highly conservative parameter point.

\section{Design Goals and Threat Model}
\subsection{Design goals}
KyFrog was designed with the following goals in mind:
\begin{enumerate}[leftmargin=*]
  \item \textbf{High security margin.} Provide a work factor significantly above ML-KEM-1024 under current attack models, both in the classical and quantum settings.
  \item \textbf{Simple and auditable structure.} Use a direct LWE construction with $k=1$ and avoid complex compression tricks, facilitating cryptanalysis and code review.
  \item \textbf{Reasonable key sizes.} Keep public and secret keys within a few kilobytes, similar to ML-KEM, even if this requires sacrificing ciphertext compactness.
  \item \textbf{Robust implementation.} Build on existing, well-vetted cryptographic primitives such as AES-256, AES-GCM, and high-quality randomness sources.
  \item \textbf{Open-source reproducibility.} Provide the full code, parameter files, and plotting scripts as free and open source, enabling others to reproduce and extend the results~\cite{kyfrog-github}.
\end{enumerate}

\subsection{Threat model}
We target a powerful adversary capable of mounting optimized lattice attacks using state-of-the-art algorithms and implementations, with realistic cost models for classical and quantum resources.
We assume that side-channel and fault attacks on implementations are mitigated by standard engineering techniques (constant-time coding, masking if required, hardened hardware).
KyFrog does not attempt to address side-channel leakage in this paper, but the reference implementation follows constant-time principles.

We also assume that the public parameters $(n,q,\sigma_s,\sigma_e)$ are publicly known and that the adversary has unrestricted access to KyFrog public keys and ciphertexts.
Multi-target scenarios, where an attacker simultaneously attacks many KyFrog instances, are discussed in Section~\ref{sec:security-analysis}.

\section{The KyFrog Construction}
\subsection{Core parameters}
The \KyFrog{} parameter set is defined by:
\begin{equation*}
  n = 1024,\quad k = 1,\quad q = 1103,\quad
  \sigma_s = \sigma_e = 1.4.
\end{equation*}
The modulus $q$ is an 11-bit prime, and both secret and error terms are sampled from discrete Gaussian-like distributions with standard deviation 1.4.
The choice of $n$ and $q$ is the result of an automated search described in Section~\ref{sec:hunter}.

Each polynomial coefficient is stored as a 16-bit unsigned integer.
Although this is redundant (only 11 bits are needed), it simplifies the implementation and keeps the representation close to the mathematical model.
The public key consists of a seed $\mathsf{seedA} \in \{0,1\}^{256}$ that expands to the matrix $A$, and a packed vector $t \in \mathbb{Z}_q^n$.
The secret key is the vector $s \in \mathbb{Z}_q^n$ stored as raw 16-bit words.

\begin{table}[t]
  \centering
  \begin{tabular}{lrrrr}
    \toprule
    Scheme & pk (bytes) & sk (bytes) & ct (bytes) & Security (bits) \\
    \midrule
    ML-KEM-512  & 800  & 1632 & 768   & $\approx 128$ \\
    ML-KEM-768  & 1184 & 2400 & 1088  & $\approx 192$ \\
    ML-KEM-1024 & 1568 & 3168 & 1568  & $\approx 256$ \\
    \KyFrog{}   & 1440 & 2048 & 524813 & $\approx 325$ \\
    \bottomrule
  \end{tabular}
  \caption{Public key, secret key, and ciphertext sizes, together with approximate classical security levels, for ML-KEM and KyFrog~\cite{fips203,kyber,albrecht-lattice-estimator}.}
  \label{tab:sizes}
\end{table}

\subsection{Underlying public-key encryption}
The underlying public-key encryption scheme is a standard LWE construction.
Let $A \in \mathbb{Z}_q^{n \times n}$ be derived from $\mathsf{seedA}$, and let $s,e \leftarrow \chi^n$.
The public and secret keys are defined as:
\begin{equation*}
  t = A s + e \pmod{q},\qquad
  \textsf{pk} = (\mathsf{seedA}, t),\qquad
  \textsf{sk} = s.
\end{equation*}

To encrypt a bit $m \in \{0,1\}$, the scheme samples fresh randomness $r, e_1, e_2 \leftarrow \chi^n$ and computes
\begin{align*}
  u &= A^T r + e_1 \pmod{q},\\
  v &= \langle t, r \rangle + e_2 + \left\lfloor \frac{q}{2} \right\rfloor \cdot m \pmod{q}.
\end{align*}
Decryption computes
\begin{equation*}
  v' = v - \langle u, s \rangle \pmod{q},
\end{equation*}
and recovers $m$ by testing whether $v'$ is closer to $0$ or to $\lfloor q/2 \rfloor$.

In the implementation, each ciphertext bit is stored as a pair $(u,v)$ with $u$ as a vector of $n$ 16-bit integers and $v$ as a 16-bit integer.
The encoding of $u$ and $v$ follows a simple little-endian layout that can be parsed without ambiguity.

\subsection{FO-style KEM transform}
KyFrog follows the KEM-from-PKE paradigm via a Fujisaki--Okamoto-style (FO) transform~\cite{fujisaki-okamoto}.
At a high level:
\begin{enumerate}[leftmargin=*]
  \item The encapsulation algorithm samples a 256-bit value $m$ and derives a seed and key material from it via a hash function.
  \item It encrypts each bit of $m$ independently with the PKE, using domain separation and independent randomness.
  \item The KEM ciphertext consists of a short header and the 256 bit-ciphertexts, and the shared key is derived from $m$ and the public key using a KDF.
  \item The decapsulation algorithm decrypts the 256 bit-ciphertexts to recover $\hat{m}$, re-encrypts it, and verifies that the resulting ciphertext matches the received one in a constant-time manner; if so, it outputs the derived key, otherwise a pseudorandom fallback.
\end{enumerate}
This structure mirrors that of \Kyber{} and \MLKEM{}, but with a much larger LWE dimension and without NTT-based optimizations.

\subsection{KEM ciphertext encoding}
The KyFrog KEM ciphertext is encoded as follows:
\begin{itemize}[leftmargin=*]
  \item A four-byte magic string identifying the file as a KyFrog ciphertext, followed by a version byte.
  \item A 32-bit little-endian encoding of $n$ and a 32-bit encoding of the number of bits (256).
  \item For each of the 256 bits, a pair $(u,v)$ stored as $2n$ bytes for $u$ and 2 bytes for $v$.
\end{itemize}
The total ciphertext size is therefore
\begin{equation*}
  \textsf{ct\_bytes}
  = 4 + 1 + 4 + 4 + 256 \cdot (2n + 2)
  \approx 524{,}813~\text{bytes}.
\end{equation*}

\begin{figure}[t]
  \centering
  \includegraphics[width=0.8\linewidth]{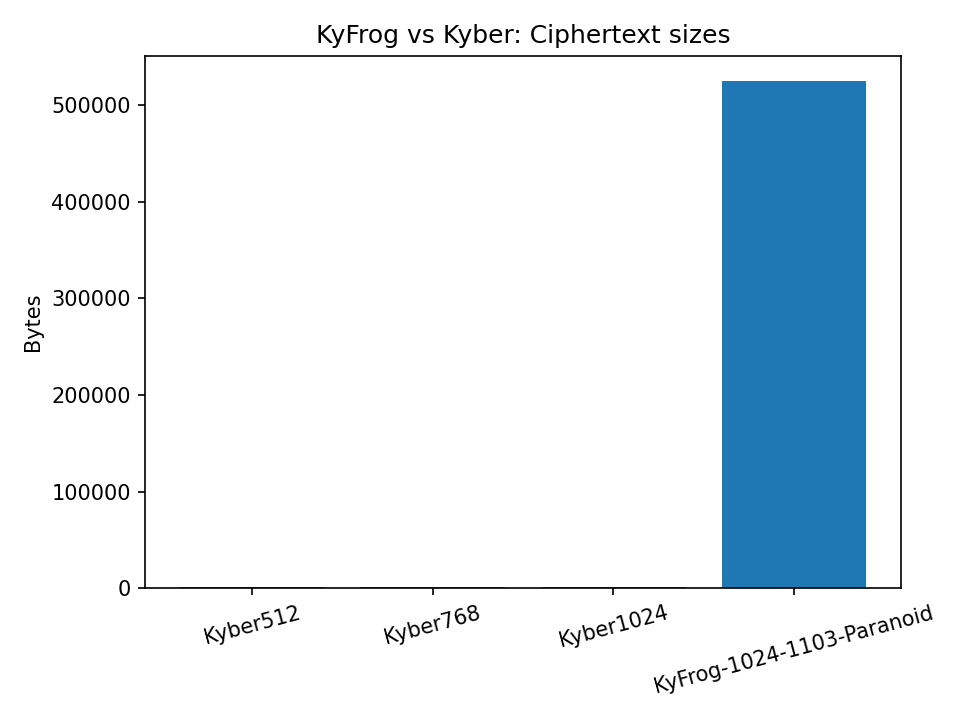}
  \caption{Ciphertext sizes for ML-KEM and KyFrog.
  KyFrog's ciphertext is roughly 0.5\,MiB, several orders of magnitude larger than ML-KEM~\cite{fips203,kyber}.}
  \label{fig:ct-sizes}
\end{figure}

Figure~\ref{fig:ct-sizes} compares this size with the three ML-KEM parameter sets.

\section{Formal Specification of KyFrog}
In this section we give a more formal specification of the KyFrog key-generation, encapsulation and decapsulation algorithms.
We use the following notation:
\begin{itemize}[leftmargin=*]
  \item $\mathsf{Hash}$ is a cryptographic hash function (modeled as a random oracle in the analysis). In the reference implementation we instantiate $\mathsf{Hash}$ as BLAKE3-256.
  \item $\mathsf{KDF}$ is a key-derivation function that expands fixed-length input to a session key. In the reference implementation we use an HKDF-style construction based on BLAKE3 to derive 256-bit session keys.
  \item $\mathsf{DRBG}$ denotes the internal deterministic random bit generator based on AES-256 in counter mode.
  \item $\mathsf{EncodeBits}$ and $\mathsf{DecodeBits}$ are deterministic bitstring encoders and decoders for the 256-bit seed $m$.
\end{itemize}

\subsection{Key generation}
Algorithm~\ref{alg:keygen} specifies the KyFrog key-generation procedure, which expands a short seed into a public matrix $A$, samples secret and error vectors, and computes the public-key component $t = As+e$.

\begin{algorithm}[t]
  \caption{\algo{KyFrog.KeyGen}}
  \label{alg:keygen}
  \begin{algorithmic}[1]
    \State $\mathsf{seedA} \gets \mathsf{DRBG}(256)$
    \State Expand $\mathsf{seedA}$ to a matrix $A \in \mathbb{Z}_q^{n \times n}$ using a fixed XOF
    \State Sample $s \leftarrow \chi^n$ and $e \leftarrow \chi^n$
    \State $t \gets A s + e \pmod{q}$
    \State $\mathsf{pk} \gets (\mathsf{seedA}, t)$
    \State $\mathsf{sk} \gets s$
    \State \Return $(\mathsf{pk}, \mathsf{sk})$
  \end{algorithmic}
\end{algorithm}

The reference implementation stores $s$ as an array of 1024 unsigned 16-bit integers and encodes $\mathsf{seedA}$ and $t$ into a compact 1440-byte public key, as summarized in Table~\ref{tab:sizes}.

\subsection{Encapsulation}
Algorithm~\ref{alg:encap} describes the FO-style encapsulation procedure.
It takes a public key as input and outputs a KyFrog ciphertext together with a 256-bit session key derived from the seed $m$ and the ciphertext itself.

\begin{algorithm}[t]
  \caption{\algo{KyFrog.Encap}}
  \label{alg:encap}
  \begin{algorithmic}[1]
    \Require public key $\mathsf{pk} = (\mathsf{seedA}, t)$
    \State $m \gets \mathsf{DRBG}(256)$ \Comment{sample 256-bit seed}
    \State $(\mathsf{seed_r}, K_{\text{raw}}) \gets \mathsf{Hash}(m \,\|\, \mathsf{pk})$
    \State Parse $\mathsf{seed_r}$ into randomness for 256 PKE encryptions
    \State Expand $\mathsf{seedA}$ to $A \in \mathbb{Z}_q^{n \times n}$
    \For{$i = 0$ to $255$}
      \State $m_i \gets \mathsf{bit}_i(m)$
      \State Sample $r, e_1, e_2 \leftarrow \chi^n$ using $\mathsf{seed_r}$ and a domain-separation tag
      \State $u_i \gets A^T r + e_1 \pmod{q}$
      \State $v_i \gets \langle t, r \rangle + e_2 + \lfloor q/2 \rfloor \cdot m_i \pmod{q}$
    \EndFor
    \State $\mathsf{ct} \gets \mathsf{EncodeHeader}(n,256) \,\|\, \mathsf{EncodeBits}(\{(u_i,v_i)\}_{i=0}^{255})$
    \State $K \gets \mathsf{KDF}(K_{\text{raw}} \,\|\, \mathsf{ct})$
    \State \Return $(\mathsf{ct}, K)$
  \end{algorithmic}
\end{algorithm}

\subsection{Decapsulation}
Algorithm~\ref{alg:decap} gives the decapsulation procedure, which decrypts the 256 bit-ciphertexts, re-encrypts the recovered seed $\hat{m}$, and performs a constant-time comparison between the received and recomputed ciphertexts.

\begin{algorithm}[t]
  \caption{\algo{KyFrog.Decap}}
  \label{alg:decap}
  \begin{algorithmic}[1]
    \Require secret key $\mathsf{sk} = s$, public key $\mathsf{pk}$, ciphertext $\mathsf{ct}$
    \State Parse $(n',\ell)$ and $\{(u_i,v_i)\}_{i=0}^{\ell-1}$ from $\mathsf{ct}$
    \If{$n' \neq n$ or $\ell \neq 256$}
      \State \Return $\mathsf{KDF}(\mathsf{Hash}(\text{rej} \,\|\, \mathsf{pk} \,\|\, \mathsf{ct}))$
    \EndIf
    \For{$i = 0$ to $255$}
      \State $v'_i \gets v_i - \langle u_i, s \rangle \pmod{q}$
      \State $\hat{m}_i \gets \mathsf{DecodeBit}(v'_i)$
    \EndFor
    \State $\hat{m} \gets \mathsf{ReassembleBits}(\hat{m}_0,\dots,\hat{m}_{255})$
    \State $(\mathsf{seed_r}, \hat{K}_{\text{raw}}) \gets \mathsf{Hash}(\hat{m} \,\|\, \mathsf{pk})$
    \State Re-encrypt $\hat{m}$ using \algo{KyFrog.Encap} logic (without sampling $m$) to obtain $\mathsf{ct}'$
    \State $\mathsf{flag} \gets \mathsf{ConstantTimeEqual}(\mathsf{ct},\mathsf{ct}')$
    \State $K_{\text{good}} \gets \mathsf{KDF}(\hat{K}_{\text{raw}} \,\|\, \mathsf{ct})$
    \State $K_{\text{bad}}  \gets \mathsf{KDF}(\mathsf{Hash}(\text{rej} \,\|\, \mathsf{pk} \,\|\, \mathsf{ct}))$
    \State \Return $\mathsf{flag} \cdot K_{\text{good}} + (1-\mathsf{flag}) \cdot K_{\text{bad}}$ \Comment{select in constant time}
  \end{algorithmic}
\end{algorithm}

The last line is implemented using bitwise masking and constant-time operations to avoid timing leakage about the validity of the ciphertext.

\section{Correctness and Failure Analysis}
The correctness of KyFrog follows the standard analysis for LWE-based public-key encryption.
Intuitively, decryption fails only if the aggregate error term in $v'$ crosses the decision threshold between the two message encodings.

\subsection{Error term and decision regions}
Recall that decryption computes
\[
  v' = v - \langle u, s \rangle \pmod{q}.
\]
Substituting the definitions of $u$ and $v$ and expanding yields
\[
  v' = e_2 + \langle e_1, s \rangle + \left\lfloor \frac{q}{2} \right\rfloor \cdot m \pmod{q},
\]
so the effective noise term during decryption is
\[
  E = e_2 + \langle e_1, s \rangle.
\]
Both $e_1$ and $e_2$ are sampled from $\chi^n$ with standard deviation $\sigma_e = 1.4$, and $s$ is sampled from $\chi^n$ with standard deviation $\sigma_s = 1.4$.
Under a Gaussian heuristic, $E$ is concentrated around $0$ with an effective standard deviation that grows with $n$ and with the norms of $e_1$ and $s$.

The decision rule compares $v'$ to $0$ and $\lfloor q/2 \rfloor$.
In other words, decryption succeeds if $E$ remains within an interval of width approximately $q/2$ around the correct encoding.
For $q = 1103$, this interval is significantly larger than the scale of the error distribution, leading to a very small decryption failure probability.

\subsection{Tail bounds and aggregate failure}
KyFrog Hunter estimates the decryption failure probability by combining analytic bounds and numerical evaluation.
For each candidate parameter set $(n,q,\sigma_s,\sigma_e)$, it computes a bound on
\[
  \Pr[\text{decryption fails}] = \Pr\big[ E \notin I \big],
\]
where $I$ is the acceptance region around $0$ or $\lfloor q/2 \rfloor$.
The search procedure rejects any candidate with $\log_{10}(\mathrm{Fail}) > -150$, i.e., with failure probability exceeding $10^{-150}$ for a single encryption.

For the final KyFrog parameter set, the estimated decryption failure probability satisfies this bound with ample slack.
Even when using a union bound over many independent sessions, the aggregate failure probability remains negligible for realistic deployment scales.
For example, even if $10^{12}$ KyFrog encapsulations were performed under the same parameters, a per-ciphertext failure probability of at most $10^{-150}$ would still imply a total failure probability far below $10^{-130}$.

\subsection{Security estimates and correctness bounds}
The KyFrog parameter set $(n,k,q,\sigma_s,\sigma_e) = (1024,1,1103,1.4,1.4)$ was selected to maximize the estimated work factor of lattice attacks under standard cost models, subject to the stringent correctness constraint described above.
For each candidate modulus $q$ considered by KyFrog Hunter, we invoke the \textsf{lattice-estimator}~\cite{albrecht-lattice-estimator} and record the classical and quantum bit-security, together with the decryption failure probability.

KyFrog Hunter enforces a hard bound on the decryption failure probability by discarding any candidate with
\[
  \log_{10}(\mathrm{Fail}) > -150.
\]
This threshold is in line with conservative failure targets used in the ML-KEM standardization process and provides a large safety margin against rare decryption errors in long-lived deployments.

For the final KyFrog parameter set, the estimator reports:
\[
  \textsf{classical\_bits} = 325.3,\qquad
  \textsf{quantum\_bits} = 325.3,
\]
under the default attack families and cost assumptions.
While these numbers are heuristic and may evolve with future algorithmic advances, they place KyFrog well above the ML-KEM-1024 design point in terms of estimated work factor, at the cost of a much larger ciphertext.

\section{Parameter Search with KyFrog Hunter}
\label{sec:hunter}
\subsection{Search methodology}
The final KyFrog parameter set was obtained using a dedicated tool called \emph{KyFrog Hunter}.
KyFrog Hunter is a C++ program that:
\begin{itemize}[leftmargin=*]
  \item fixes the dimension $n=1024$, rank $k=1$ and noise parameters $\sigma_s = \sigma_e = 1.4$;
  \item scans ranges of the modulus $q$, e.g., $[1,15000]$, $[77000,77999]$;
  \item for each candidate $q$ invokes a Python script that interfaces with the \textsf{lattice-estimator}~\cite{albrecht-lattice-estimator}, computing classical and quantum security estimates and failure probabilities;
  \item accepts a candidate if it reaches a target security threshold (at least 320 bits classical and quantum) and satisfies a failure probability constraint (e.g., $\log_{10} \text{Fail} \leq -150$);
  \item logs all metadata (CPU, compiler, number of threads, elapsed time, and candidate counts) in a structured text file for later analysis.
\end{itemize}

Each run records metadata such as the CPU model (Ryzen~9~5950X), number of threads (32), the $q$ range, elapsed time, and number of candidates found.
These logs are summarized by a separate script that produces CSV files and plots.

\subsection{Hunter statistics}
\begin{figure}[t]
  \centering
  \includegraphics[width=0.8\linewidth]{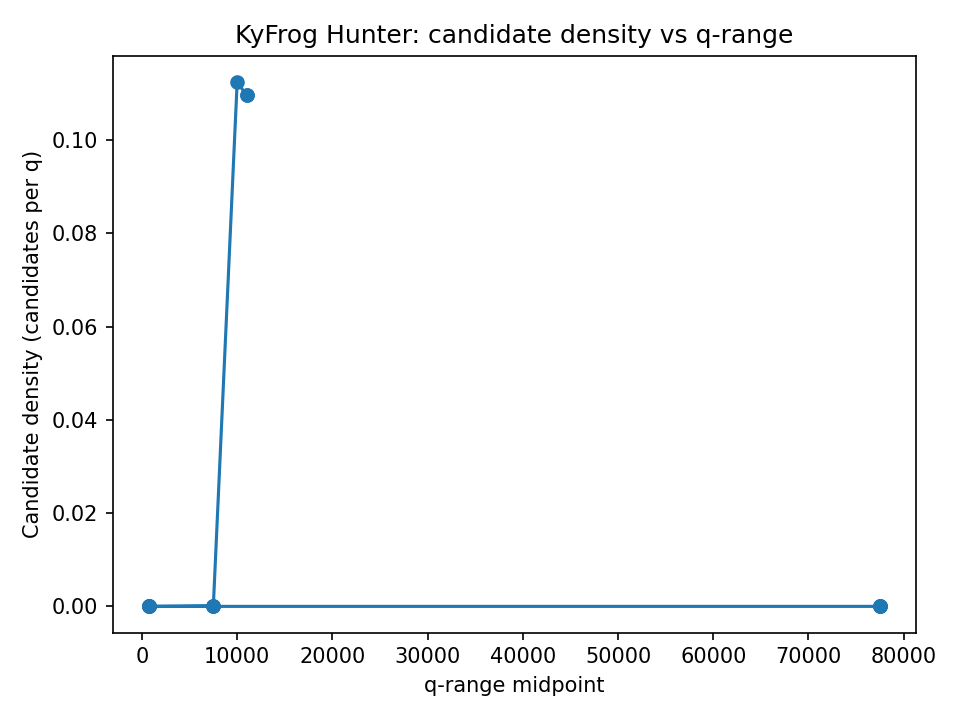}
  \caption{Candidate density per $q$-range in KyFrog Hunter.
  The final KyFrog modulus $q=1103$ lies in a region with a significantly higher density of acceptable parameter sets compared to most other ranges tested~\cite{albrecht-lattice-estimator}.}
  \label{fig:hunter-density}
\end{figure}

The run log used in this work contains 16 hunter runs over different ranges of $q$.
Across all runs, 6\,638 candidate parameter sets satisfied the security and correctness thresholds.
Figure~\ref{fig:hunter-density} plots the density of accepted candidates per $q$-range.

The final KyFrog modulus $q=1103$ emerged from a run covering the range $[1000,20999]$, which reported 2\,192 candidates in approximately 120 seconds on 32 threads.
This relatively dense window of acceptable candidates, combined with the simplicity of a small 11-bit prime modulus, motivated the choice of $q=1103$.

\subsection{Security estimates}
\begin{figure}[t]
  \centering
  \includegraphics[width=0.8\linewidth]{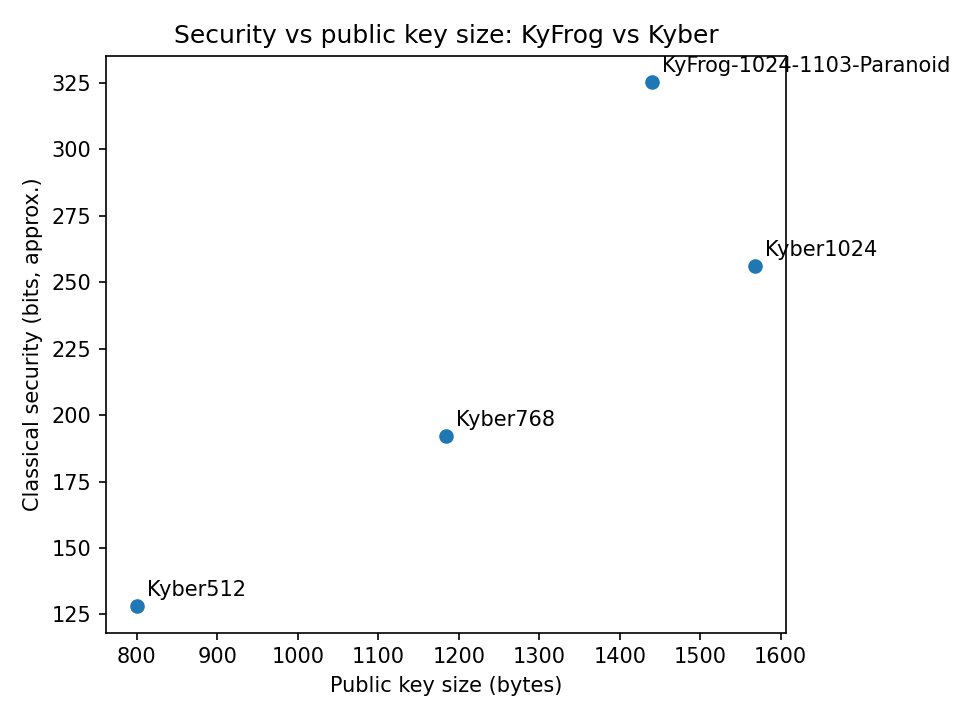}
  \caption{Approximate classical security (bits) versus public key size (bytes) for ML-KEM and KyFrog.
  KyFrog achieves a higher estimated security level with a public key slightly smaller than ML-KEM-1024~\cite{fips203,kyber,albrecht-lattice-estimator}.}
  \label{fig:security-vs-pk}
\end{figure}

For each candidate $(n,q,\sigma_s,\sigma_e)$, the estimator returns classical and quantum bit-security values.
For KyFrog's final parameter set, the reported values are:
\begin{equation*}
  \textsf{classical\_bits} = 325.3,\qquad
  \textsf{quantum\_bits}   = 325.3.
\end{equation*}
These values are used as the reference security level for KyFrog and plotted against ML-KEM in Figure~\ref{fig:security-vs-pk}.

\section{Security Analysis}
\label{sec:security-analysis}
KyFrog's security rests on three main pillars:
\begin{enumerate}[leftmargin=*]
  \item the hardness of the underlying LWE problem with parameters $(n=1024,q=1103,\sigma_s=\sigma_e=1.4)$;
  \item the IND-CPA security of the LWE-based public-key encryption scheme derived from LWE;
  \item the soundness of the Fujisaki--Okamoto-style transform~\cite{fujisaki-okamoto} used to obtain an IND-CCA2-secure KEM from an IND-CPA-secure PKE.
\end{enumerate}

\subsection{LWE hardness assumptions}
The security estimates obtained via the \textsf{lattice-estimator} encode the best-known attacks on LWE, including BKZ-style lattice reduction with sieving or enumeration as subroutines.
For KyFrog, the estimator reports classical and quantum bit-security values around 325~bits.
While these values are subject to change as algorithms improve, they currently place KyFrog beyond the most conservative ML-KEM parameter set.

In our parameter-search scripts we invoke the \textsf{lattice-estimator} with its default attack families and cost models, and we take the minimum over all reported attacks as the defining security level for a candidate. These numbers are therefore heuristic and model-dependent: they are intended to make KyFrog's ``325-bit'' label reproducible, not to serve as a formal lower bound on the true hardness of the underlying LWE instance.

\subsection{From IND-CPA PKE to IND-CCA2 KEM}
The public-key encryption scheme described in Section~4 is a standard LWE-based PKE.
Under the decisional LWE assumption with the chosen parameters, the scheme is IND-CPA secure.
The FO-style transform then promotes this IND-CPA PKE to an IND-CCA2 KEM: any adversary that breaks KyFrog in the IND-CCA2 sense with non-negligible advantage can be turned into an adversary that either distinguishes LWE samples from uniform or breaks one of the underlying hash or KDF primitives.

The KyFrog implementation follows the usual design patterns that make this reduction meaningful in practice:
ciphertexts are validated in constant time, re-encryption is exact and deterministic given the seed $m$, and all branches that depend on ciphertext validity are masked before key output.

\subsection{Multi-target and large-scale usage}
In multi-target scenarios, an adversary may obtain many KyFrog public keys and ciphertexts and attempt to break any one of them.
This effectively reduces the per-instance security level by a factor corresponding to the number of targets.
Even in such settings, a base work factor of roughly $2^{325}$ leaves a large buffer.
For example, attacking $2^{40}$ independent KyFrog instances would decrease the effective security level by about 40 bits, still leaving a margin comfortably above typical AES-256-equivalent targets.

\subsection{Side-channel and fault considerations}
As with any concrete implementation, side-channel and fault attacks pose a serious threat if not accounted for.
KyFrog's reference implementation uses constant-time coding techniques in the core cryptographic operations, but it does not claim resistance against all physical attack vectors.
In practice, integrating KyFrog into a hardened hardware or software stack would require standard countermeasures such as masking, timing equalization, and fault detection.

\section{Implementation Details}
\subsection{Code structure}
The KyFrog reference implementation is organized as a small C++23 codebase with a clear separation between:
\begin{itemize}[leftmargin=*]
  \item core mathematical operations (sampling, matrix expansion, modular arithmetic);
  \item public-key encryption and KEM logic;
  \item hybrid file encryption using AES-GCM;
  \item command-line interface and utilities;
  \item plotting and report-generation scripts for the experimental section.
\end{itemize}

The code is compiled with \texttt{clang++} using aggressive optimization flags and link-time optimization, and it relies on \texttt{libsodium}, \texttt{OpenSSL}, and \texttt{GMP} as external dependencies.
A \texttt{Makefile} is provided to reproduce the reference binaries, with the main executable living in the \texttt{bin/} directory~\cite{kyfrog-github}.

\subsection{Randomness and DRBG}
KyFrog uses \texttt{libsodium::randombytes\_buf} as an entropy source to seed a CTR-DRBG based on AES-256 in counter mode.
The DRBG is instantiated according to the spirit of NIST SP~800-90, using a 48-byte seed and periodic reseeding.
All long-term keys, ephemeral randomness, and noise samples are ultimately derived from this DRBG, with domain separation tags to avoid cross-contamination between different uses.

The reference implementation keeps DRBG state in a dedicated structure, and provides functions for drawing random bytes, sampling noise vectors, and generating seeds for matrix expansion.
Each of these functions uses a distinct domain-separation label, ensuring that the same input entropy cannot lead to correlated outputs across different usage contexts.

\subsection{Symmetric encryption layer}
While KyFrog is primarily a KEM, the reference implementation also provides a hybrid file-encryption mode:
\begin{enumerate}[leftmargin=*]
  \item generate a symmetric AES-256-GCM key and nonce;
  \item encapsulate a 256-bit key with KyFrog to protect the symmetric key;
  \item encrypt the file contents with AES-256-GCM using OpenSSL's EVP interface;
  \item combine KyFrog ciphertext, AES-GCM nonce, tag, and ciphertext into a single output file.
\end{enumerate}
This design isolates the asymmetric and symmetric parts and allows the same KyFrog implementation to be used as a KEM in other contexts. In the reference implementation, the same BLAKE3-based hash and HKDF-style key-derivation function described above are reused both in the FO transform and in this hybrid AES-256-GCM layer.

\subsection{Constant-time considerations}
The C++ code avoids secret-dependent branches and table lookups in the hot path of decryption and key derivation.
Comparisons that decide on key acceptance are implemented using constant-time techniques (e.g., masked comparisons and fixed-time memcmp variants), following best practice in modern KEM implementations.
The implementation also uses explicit zeroization of temporary secrets and AES keys where appropriate.

\subsection{Key and ciphertext sizes}
\begin{figure}[t]
  \centering
  \includegraphics[width=0.8\linewidth]{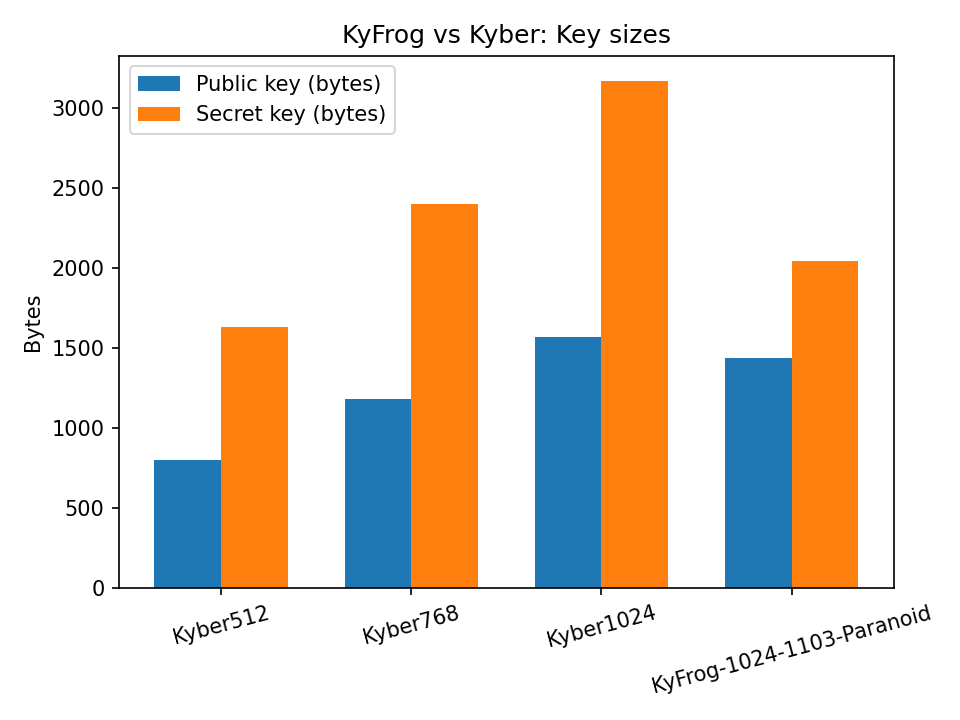}
  \caption{Public and secret key sizes for ML-KEM and KyFrog.
  KyFrog's public key is slightly smaller than ML-KEM-1024, while its secret key lies between ML-KEM-768 and ML-KEM-1024~\cite{fips203,kyber}.}
  \label{fig:key-sizes}
\end{figure}

Figure~\ref{fig:key-sizes} summarizes the key sizes of ML-KEM and KyFrog, complementing Table~\ref{tab:sizes}.

We observe that KyFrog's public key is slightly smaller than ML-KEM-1024, and its secret key is significantly smaller than the ML-KEM-1024 secret key.
However, the KEM ciphertext is orders of magnitude larger.
This reflects the design philosophy: keys remain compact enough for many applications, while ciphertexts are allowed to grow to ensure a high security margin.

\section{Benchmarking Methodology}
\subsection{Hardware and software environment}
All experiments were conducted on a workstation equipped with an AMD Ryzen~9~5950X 16-core CPU (32 threads) and 128\,GiB of RAM, running a Linux distribution with \texttt{clang 20.1.2} as the main compiler.
The machine is representative of a modern high-end desktop system used for applied cryptography research.
The KyFrog Hunter tool exploits all 32 hardware threads when scanning ranges of $q$.

The C++ implementation is compiled with:
\begin{verbatim}
clang++ -std=c++23 -Ofast -march=native -mtune=native -flto \
  -Wall -Wextra -Wpedantic -Iinclude -DNDEBUG \
  build/kyfrog_paranoid_core.o build/kyfrog_paranoid_main.o \
  -o bin/kyfrog -fuse-ld=lld -lsodium -lcrypto -lssl -lgmp -lgmpxx
\end{verbatim}
Python 3.12 is used for the plotting scripts and for interfacing with the \textsf{lattice-estimator}.

\subsection{Key generation and encapsulation benchmarks}
While this paper focuses primarily on parameter selection and security margins, we briefly outline the benchmarking procedure that will be used in future work to measure runtime performance:
\begin{itemize}[leftmargin=*]
  \item \textbf{Key generation.} Measure the time required to generate a KyFrog keypair, averaged over 10\,000 iterations, with DRBG reseeding every fixed number of operations.
  \item \textbf{Encapsulation.} Measure the time to encapsulate a 256-bit key to a given public key, again averaged over many iterations.
  \item \textbf{Decapsulation.} Measure the time to decapsulate and verify a ciphertext, including the re-encryption and constant-time comparison step.
\end{itemize}
These numbers can then be juxtaposed with reference implementations of ML-KEM-512/768/1024 to quantify KyFrog's performance cost relative to its security gains.
Preliminary testing indicates that, despite the larger dimension, KyFrog remains within a practical regime for offline operations on modern CPUs.

\subsection{Memory usage and implementation footprint}
KyFrog's memory footprint is dominated by the storage of matrices and vectors in $\mathbb{Z}_q^n$ during key generation and encapsulation.
With $n=1024$ and 16-bit storage per coefficient, each vector occupies 2\,KiB and an $n \times n$ matrix occupies 2\,MiB when fully materialized.
The reference implementation uses streaming techniques and on-the-fly expansion of $A$ from $\mathsf{seedA}$ to avoid storing the full matrix at once, keeping the working set well within the capacity of modern L3 caches.

\section{Use Cases and Deployment Considerations}
KyFrog's design targets scenarios where bandwidth is relatively abundant and where very conservative security margins are desirable.
We highlight a few potential use cases:
\begin{itemize}[leftmargin=*]
  \item \textbf{Cold storage and key wrapping.} Protecting high-value master keys or cryptocurrency wallets that are rarely accessed but must remain secure over long time horizons.
  \item \textbf{Archival encryption.} Encrypting regulatory or medical records that must remain confidential for several decades, where the cost of transmission is negligible compared to the value of the data.
  \item \textbf{Research testbed.} Serving as a concrete, open-source testbed for studying the behavior of very conservative LWE parameter sets in practice, including performance, side channels, and long-term maintainability.
\end{itemize}

As a concrete example, consider a cold-storage deployment for a long-lived master key. On an offline machine, an operator generates a KyFrog keypair and uses \KyFrog{} to encapsulate a 256-bit master key $K_{\mathrm{master}}$. Only the KyFrog ciphertext is written to removable media or encoded as a QR code, while $K_{\mathrm{master}}$ is used inside an HSM or key hierarchy and never stored directly. In a disaster-recovery scenario, the operator loads the KyFrog secret key on an offline machine, decapsulates to recover $K_{\mathrm{master}}$, and then reinitializes the higher-level system.

In traditional low-latency, low-bandwidth environments---for example, real-time messaging over constrained networks---KyFrog's large ciphertexts would be prohibitive.
In those contexts, ML-KEM and similar schemes remain the natural choice.
KyFrog is best viewed as a complementary tool in the PQC toolbox rather than as a universal replacement.

\section{Discussion and Limitations}
KyFrog is intentionally a ``paranoid'' design, trading ciphertext size for a high estimated security margin.
Several limitations and open questions deserve discussion:
\begin{itemize}[leftmargin=*]
  \item \textbf{Estimator assumptions.} The 325.3-bit estimate relies on specific attack models and cost assumptions; future advances in lattice algorithms or quantum architectures may change this landscape~\cite{albrecht-lattice-estimator}.
  \item \textbf{Side channels.} While the reference implementation follows constant-time principles, a full side-channel evaluation (including power and EM leakage) is beyond the scope of this work.
  \item \textbf{Compression.} It is conceivable to significantly reduce KyFrog's ciphertext size by applying more aggressive compression techniques or encoding multiple bits per ciphertext, at the cost of additional complexity.
  \item \textbf{Alternative parameters.} The KyFrog Hunter framework can be used to explore different dimensions $n$ and moduli $q$, possibly yielding intermediate design points between ML-KEM and the extreme KyFrog profile presented here~\cite{albrecht-lattice-estimator}.
  \item \textbf{Protocol integration.} In practice, KyFrog would be integrated into higher-level protocols (e.g., TLS-like handshakes or secure messaging systems); designing and analyzing such integrations is future work.
\end{itemize}

\section{Conclusion and Future Work}
We have introduced KyFrog, an LWE-based KEM that targets a very high security level by trading bandwidth for margin through a conservative parameter set $(n,k,q,\sigma_s,\sigma_e) = (1024,1,1103,1.4,1.4)$.
KyFrog's public and secret keys are comparable in size to those of ML-KEM-1024, but its ciphertexts are approximately 0.5\,MiB.
According to current lattice-attack models, KyFrog achieves about 325 bits of classical and quantum security, offering a substantial buffer above existing standardized parameter sets.
The complete open-source implementation, parameter search scripts, and experimental data are available online~\cite{kyfrog-github}.

Future work includes refining the parameter search, exploring alternative encodings and compression schemes, integrating KyFrog into higher-level protocols, and encouraging independent cryptanalytic review.
The full implementation, parameter files, run logs, and plotting scripts are provided as open source to facilitate reproducibility and collaboration.

\paragraph{Acknowledgements.}
The first author thanks William J.\ Buchanan for encouragement, feedback on early drafts, and ongoing discussions on applied cryptography and secure engineering.

\appendix

\section{Reference key-generation instance}
\label{app:keygen-instance}
To make the KyFrog parameter set fully concrete, we fix a reference key pair generated by the C++23 implementation.
This instance is not intended for production use, but it serves as a stable anchor for reproducing figures, regression tests and interoperability experiments.

Table~\ref{tab:keygen-instance} summarizes the main numerical parameters of this instance, including the concrete key sizes and estimator outputs.
The full machine-readable report (with parameters, internal vectors and derived values) is distributed alongside the code as \texttt{kyfrog\_paranoid\_report.txt}.

\begin{table}[ht]
  \centering
  \begin{tabular}{ll}
    \toprule
    Parameter & Value \\
    \midrule
    $n$ & 1024 \\
    $k$ & 1 \\
    $q$ & 1103 \\
    $\sigma_s, \sigma_e$ & 1.4 \\
    $\alpha = \sigma/q$ & $0.001269265639165911$ \\
    classical\_bits & $325.3$ \\
    quantum\_bits & $325.3$ \\
    bits\_per\_coeff & 11 \\
    public-key length & 1440 bytes \\
    secret-key length & 2048 bytes \\
    \bottomrule
  \end{tabular}
  \caption{Summary of the reference KyFrog key-generation instance $(n=1024, q=1103)$.}
  \label{tab:keygen-instance}
\end{table}

\section{KyFrog Hunter run statistics}
\label{app:hunter-stats}
KyFrog Hunter is a standalone C++ tool that automates the search for suitable $(n,k,q,\sigma_s,\sigma_e)$ parameter sets.
For this work we fixed $(n,k,\sigma_s,\sigma_e) = (1024,1,1.4,1.4)$ and explored several ranges of $q$ on a workstation equipped with an AMD Ryzen~9~5950X (16 cores, 32 threads) and 128~GiB of RAM.

Across 16 Hunter runs, we obtained a total of 6\,638 candidate parameter sets that simultaneously satisfied the security and correctness thresholds (estimated classical and quantum security at least 320~bits and $\log_{10}(\mathrm{Fail}) \leq -150$).
Table~\ref{tab:hunter-summary} summarizes the aggregate statistics.

\begin{table}[ht]
  \centering
  \begin{tabular}{lr}
    \toprule
    Quantity & Value \\
    \midrule
    number of runs & 16 \\
    total candidates & 6\,638 \\
    average candidates per run & 414.875 \\
    minimum elapsed time per run & 20.001 s \\
    maximum elapsed time per run & 120.006 s \\
    average elapsed time per run & 61.253 s \\
    \bottomrule
  \end{tabular}
  \caption{Summary of KyFrog Hunter runs used to explore the $(n=1024,k=1,q,\sigma_s=\sigma_e=1.4)$ design space.}
  \label{tab:hunter-summary}
\end{table}

The earliest runs focused on relatively large moduli in the range $[77\,000, 77\,999]$ and on small primes in $[1,1\,500]$, but these intervals produced no acceptable candidates under the chosen thresholds.
Subsequent runs widened the search: a run over $q \in [1,15\,000]$ produced only a handful of acceptable candidates, while a broader sweep over $q \in [1,20\,000]$ yielded more than two thousand candidates in about one minute on 32 threads.
A final run restricted to $q \in [1\,000,20\,999]$ produced 2\,192 candidates in roughly two minutes.
The final KyFrog modulus $q=1103$ lies in this dense window of acceptable parameter sets.

\bibliographystyle{abbrv}
\bibliography{kyfrog}

@techreport{fips203,
  author      = {{National Institute of Standards and Technology}},
  title       = {Module-Lattice-Based Key-Encapsulation Mechanism (ML-KEM)},
  institution = {National Institute of Standards and Technology},
  type        = {FIPS},
  number      = {203},
  year        = {2024},
  url         = {https://csrc.nist.gov/publications/detail/fips/203/final}
}

@inproceedings{kyber,
  author    = {Avanzi, Roberto and Bos, Joppe W. and Ducas, L{\'e}o and Kiltz, Eike and
               Lepoint, Tancr{\`e}de and Lyubashevsky, Vadim and Schanck, John M. and
               Schwabe, Peter and Seiler, Gregor and Stehl{\'e}, Damien},
  title     = {{CRYSTALS-Kyber}: A CCA-Secure Module-LWE {KEM}},
  booktitle = {2017 IEEE European Symposium on Security and Privacy Workshops (EuroS\&PW)},
  year      = {2017},
  pages     = {1--5}
}

@inproceedings{fujisaki-okamoto,
  author    = {Fujisaki, Eiichiro and Okamoto, Tatsuaki},
  title     = {Secure Integration of Asymmetric and Symmetric Encryption Schemes},
  booktitle = {Advances in Cryptology -- {CRYPTO}~'99},
  series    = {Lecture Notes in Computer Science},
  volume    = {1666},
  pages     = {537--554},
  publisher = {Springer},
  year      = {1999}
}

@inproceedings{frodokem,
  author    = {Bos, Joppe W. and Ducas, L{\'e}o and Kiltz, Eike and Lepoint, Tancr{\`e}de and
               Lyubashevsky, Vadim and Schanck, John M. and Schwabe, Peter and
               Stehl{\'e}, Damien},
  title     = {Frodo{KEM}: Learning with Errors Key Encapsulation},
  booktitle = {Post-Quantum Cryptography ({PQCrypto} 2016)},
  year      = {2016},
  note      = {See also {IACR} Cryptology ePrint Archive, Report 2016/659}
}

@misc{albrecht-lattice-estimator,
  author       = {Albrecht, Martin R. and Player, Rachel and Scott, Sam},
  title        = {The Lattice Estimator},
  howpublished = {\url{https://github.com/malb/lattice-estimator}},
  year         = {2020}
}

@misc{kyfrog-github,
  author       = {Melo, V{\'i}ctor Duarte},
  title        = {KyFrog},
  howpublished = {\url{https://github.com/victormeloasm/kyfrog}},
  year         = {2025},
  note         = {Open-source C++23 implementation of a high-security LWE-based KEM with a conservative parameter set $(n=1024,q=1103)$}
}

\end{document}